\documentclass[journal]{IEEEtranTIE}
\usepackage{graphicx}
\usepackage{cite}
\usepackage{picinpar}
\usepackage{amsmath}
\usepackage{amsfonts}
\usepackage{url}
\usepackage[utf8]{inputenc}
\usepackage{colortbl}
\usepackage{soul}
\usepackage{multirow}
\usepackage{pifont}
\usepackage{color}
\usepackage{alltt}
\usepackage[hidelinks]{hyperref}
\usepackage{enumerate}
\usepackage{siunitx}
\usepackage{breakurl}
\usepackage{epstopdf}
\usepackage{pbox}
\usepackage[scaled]{helvet}     
\graphicspath{{./figure/}}
\usepackage{mathtools}
 
\usepackage{svg}
\usepackage{bm}
\usepackage{booktabs}
\usepackage{makecell}
\usepackage{array}
\usepackage{placeins}      
\usepackage{booktabs}
\usepackage{multirow}
\usepackage[flushleft]{threeparttable}
\usepackage{makecell}

\usepackage{caption}
\usepackage[labelformat=simple]{subcaption}


\renewcommand{\tablename}{Table }
\renewcommand{\figurename}[1]{Fig. #1}

\begin{document}
\title{DSP-Based Sub-Switching-Period Current-Limiting Control
      for Grid-Tied Inverter under Grid Faults}
\author{
	\vskip 1em
	
	Jaeyeon Park,
	Jiyu Lee,
	Junyeol Maeng, \emph{Graduate Student Member, IEEE}, and Shenghui Cui, \emph{Member, IEEE}

	\thanks{
J. Park, J. Lee, J. Maeng, and S. Cui 
are with the Department of Electrical and Computer Engineering
and SNU Electric Power Research Institute,
Seoul National University, Seoul 08826, Republic of Korea 
(e-mail: ok6530@snu.ac.kr; jiyu0805@snu.ac.kr; junyul7@snu.ac.kr; cuish@snu.ac.kr)
	}
}
 
\maketitle

\markboth{IEEE TRANSACTIONS ON INDUSTRIAL ELECTRONICS}%
{}

\definecolor{limegreen}{rgb}{0.2, 0.8, 0.2}
\definecolor{forestgreen}{rgb}{0.13, 0.55, 0.13}
\definecolor{greenhtml}{rgb}{0.0, 0.5, 0.0}

\begin{abstract}
    This paper presents a sub-switching period current-limiting control for a grid-tied inverter 
    to prevent transient overcurrents during grid faults and enable seamless fault ride-through (FRT). 
    Sudden grid-voltage disturbances, such as voltage sags or phase jumps, can induce large transient currents within a switching period, 
    particularly at low switching frequencies.
    Upon disturbance detection, the proposed method immediately modifies the pulse-width modulation carrier, 
    enabling continuous regulation of the inverter output current within a time much shorter than a switching period without interrupting current flow. 
    The proposed method can be implemented on commonly used digital signal processors without requiring specialized analog or digital circuits or high-speed computing devices. 
    Experimental results
    from a 2-level, 3-phase inverter 
    switching at 3.6 kHz validate the effectiveness of the proposed method
    under symmetric and asymmetric voltage sags and phase jumps.
\end{abstract}

\section{Introduction} \label{sec:introduction}

\IEEEPARstart{W}{ith} the increase of the inverter-based resources
in power grids, grid-tied inverters are required to have
fault ride-through (FRT) capability to ensure power system reliability \cite{IEEESTD1547_2018,ramirez2024review}.
At the instant of grid faults, such as voltage sags and phase jumps,
large transient currents can be induced in grid-tied inverters, 
particularly when operating at low switching frequencies
and with low filter inductance,
e.g., in MW-scale high-power inverters
\cite{Nagai2018ZVRTCapability,Li2022AnInrushCurrent}.
Since power semiconductor devices are vulnerable to overcurrent stress \cite{Wu2015AComprehensiveInvestigation},
large transient currents may trigger the protective tripping of inverters,
potentially causing power system instability due to the sudden disconnection of generation sources\cite{Lara2025April28th2025}.
Therefore, 
mitigating overcurrent at the instant of grid faults is crucial
to ensure continuous operation without tripping events
and provide the stable FRT capability of grid-tied inverters. 
This paper focuses on mitigating instantaneous overcurrent within the switching period right after a fault occurrence.
%
 
\begin{figure}[t!]
    \centering
    \includegraphics[width=1.0\linewidth]{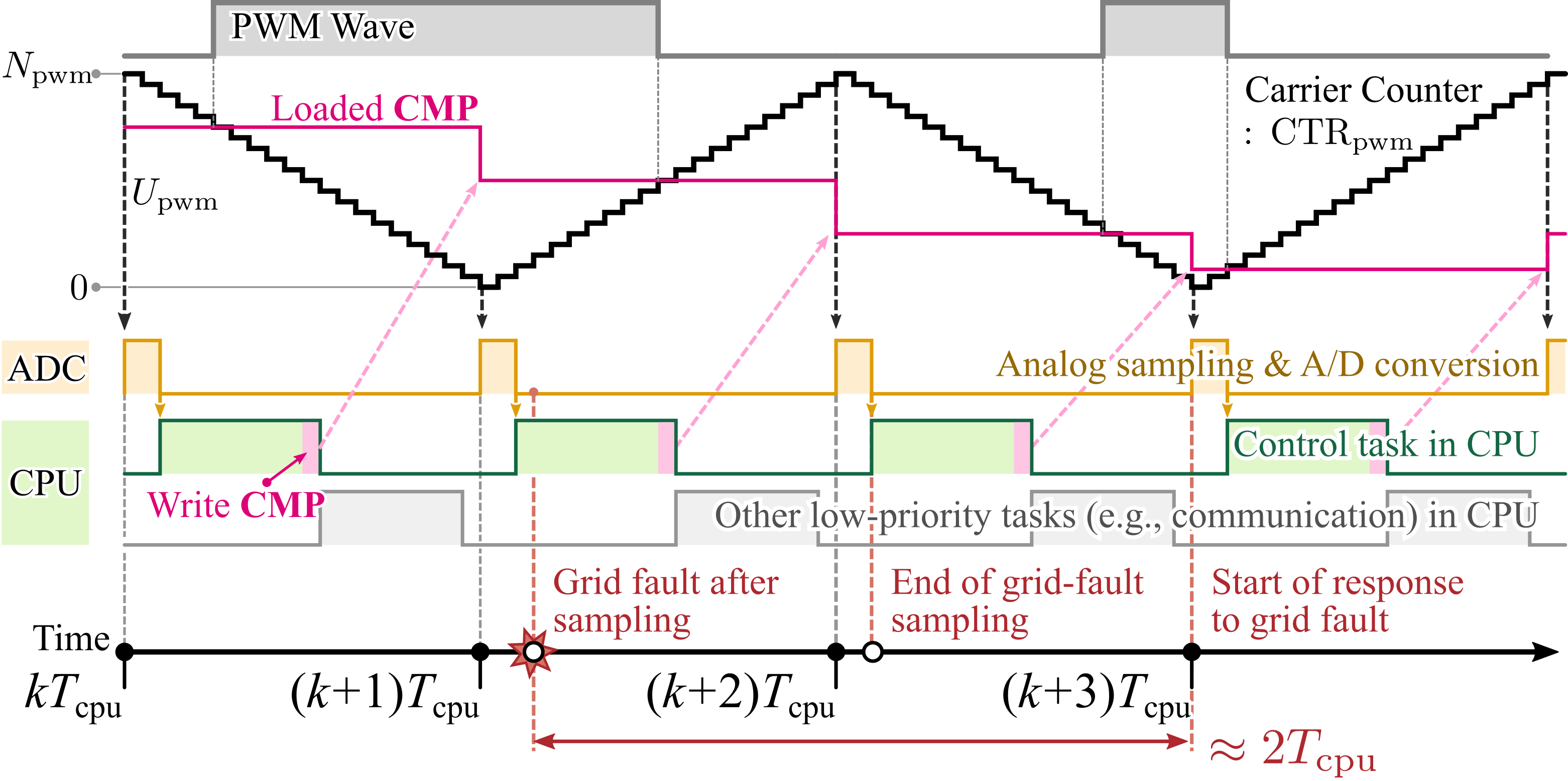}
    \caption{Worst-case grid-fault scenario in a double-sampling, double-update control structure.
    `CTR' and `CMP' denote the carrier counter and the compare value for PWM generation, respectively.}
    \label{fig:dsdu_worstcase}
    \vspace{-1.5em}
\end{figure}

These overcurrent transients
are mainly attributed to the control delay
inherent in conventional digital control systems \cite{Li2016InrushTransientCurrent}. 
These systems typically adopt 
single-sampling, single-update (SSSU) or double-sampling, double-update (DSDU) structures\cite{He2022ReviewOfMultisampling}.
In these structures, once the duty cycle is determined for a control period,
the controller cannot respond to grid faults occurring within that timeframe.
As illustrated in \figurename{\ref{fig:dsdu_worstcase}},
in the DSDU structure, the controller cannot respond to a fault occurring immediately after the sampling instant
for nearly two digital control periods, $2T_{\mathrm{cpu}}$, leading to an uncontrolled current surge.
Considering 
the equivalent zero-order holder (ZOH) nature of the
pulse-width modulation (PWM) synthesis, the delay can be as long as $2.5T_{\mathrm{cpu}}$ \cite{Li2022AnInrushCurrent}.

To address this issue, 
several studies have proposed methods to reduce the response delay to grid faults.
First,
hardware-based methods capable of rapid fault detection have been investigated \cite{Liu2009Triple, Nagai2018ZVRTCapability,Li2022AnInrushCurrent,Wu2025ATransientCurrentLimiting}. 
In these studies, transient overcurrents are suppressed 
by temporarily turning off the switches upon detecting a voltage sag. 
To achieve high-speed response, analog circuits and dedicated hardware, such as field-programmable gate array (FPGA), are employed.
In \cite{Mohiuddin2025ATwoStageCurrent}, 
this approach has been combined with other current limiting strategies,
such as virtual impedance control\cite{Paquette2015VirtualImpedanceCurrent}, as a primary protection stage for high-speed current limiting.
However, despite ensuring rapid response to grid faults,
these methods require specialized hardware, increasing system complexity and cost.
Furthermore, blocking the gate signals disrupts the voltage reference synthesis, potentially
causing integrator windup in the upper-level current controllers \cite{Baeckeland2024OvercurrentLimitinginGridForming}.

Alternatively, 
control-based methods 
have been proposed 
to mitigate transient fault currents without additional hardware and gate signal blocking
\cite{Dong2023APLL_lessVoltage, Kawashima2024UltraRobust1MHz}.
In \cite{Dong2023APLL_lessVoltage},
a deadbeat current controller 
is adopted to regulate the inverter output current and 
predict grid voltage disturbances in the SSSU control structure.
While not as instantaneous as hardware-based approaches, 
it enables 
the digital controller to respond to grid faults after one switching period.
In \cite{Kawashima2024UltraRobust1MHz},
a multisampling technique combined with double-update PWM is developed based on FPGA, 
and it reduces the upper bound of the delay
of reaction to grid faults to a half of the switching period.
However, these control-based approaches still face limitations
in achieving instantaneous current limiting comparable to hardware-based approaches,
and often require high-performance computing processors.

To overcome these limitations,
this paper proposes a simple current-limiting strategy 
that can be implemented on commonly used digital signal processors (DSPs)
without specialized hardware.
By modifying the PWM carrier counter upon fault detection,
the proposed method enables grid-tied inverters to respond to grid faults immediately within a switching period 
and continuously regulate the output current
without interrupting current flow.

The rest of this paper is organized as follows.
Section II describes the system configuration 
and the proposed current-limiting strategy.
Section III presents experimental results under various grid faults including symmetric
and asymmetric voltage sags and phase jumps.
Finally, Section IV concludes the paper.


\section{Proposed Current-Limiting Control} \label{sec:proposed_current_limiting_control}
\begin{figure}[t]
    \centering
    \includegraphics[width=1.0\linewidth]{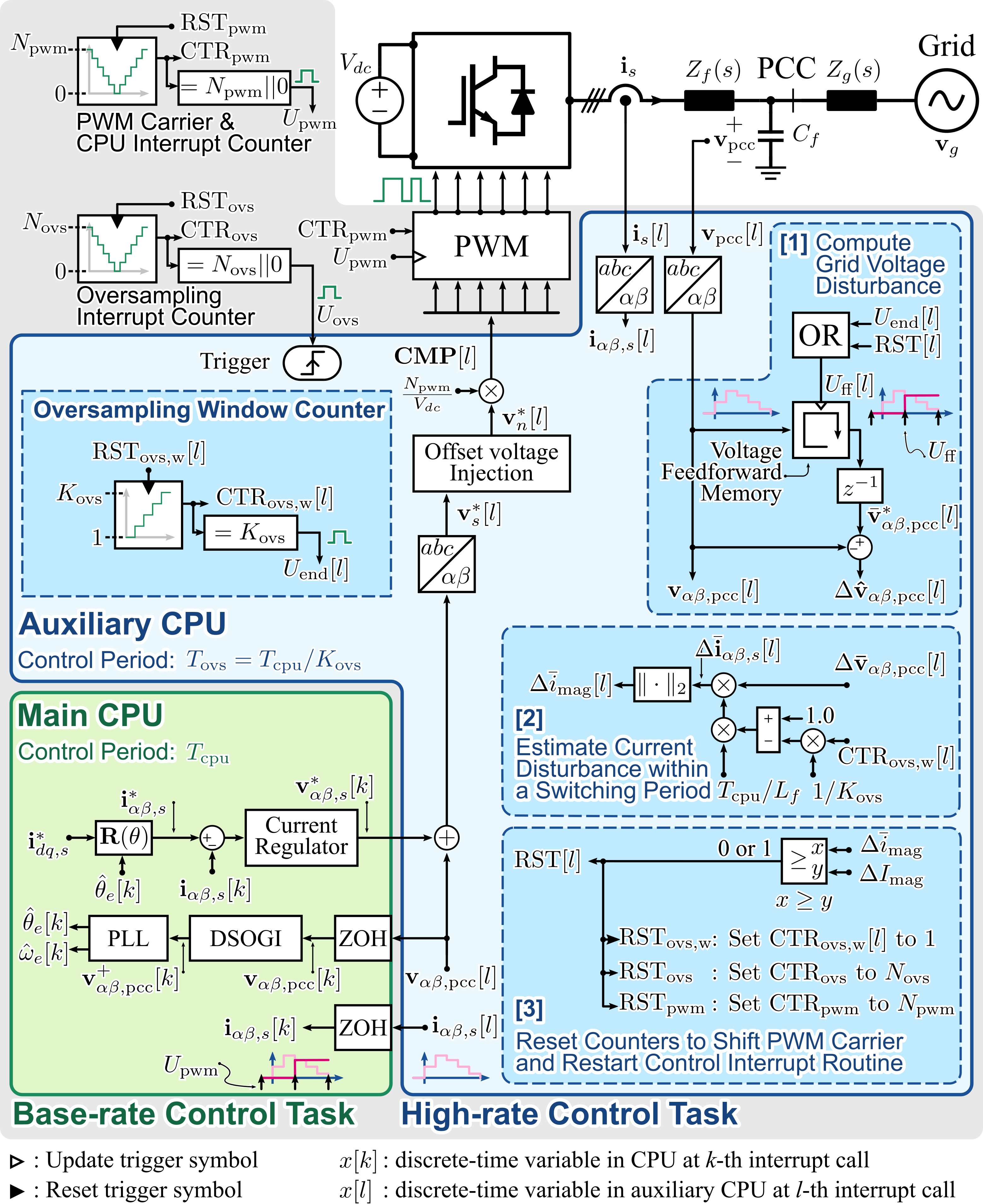}
    \caption{Proposed control structure of a grid-tied inverter 
    using multisampling technique and instantaneous PWM carrier shift when a grid fault occurs.}
    \label{fig:proposed_control_loop}
    \vspace{-1em}
\end{figure}

This section presents the proposed control structure and details the operating principles of the current-limiting strategy.
First, the overall multirate control architecture is introduced. 
Then, the method for detecting grid voltage disturbances and 
the subsequent modification of the PWM carrier and control routines
to suppress transient overcurrents are explained.

\subsection{Overview of Control Structure and Normal Operation} \label{sec:overview_of_the_proposed_control_structure}

\begin{figure}[!t]
    \centering
    \includegraphics[width=1.0\linewidth]{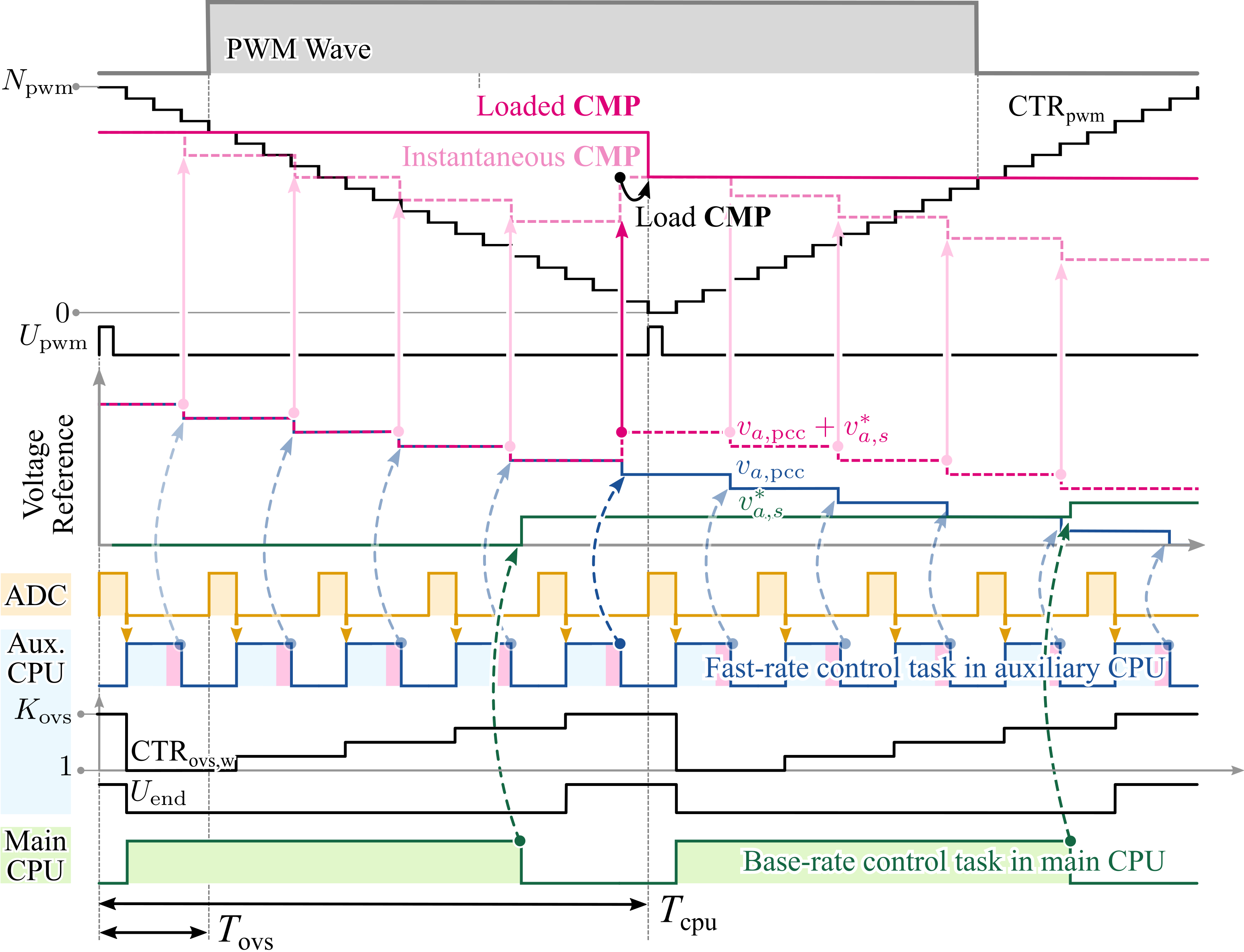}
    \caption{PWM update timing diagram of the proposed control structure during the normal operation when $K_{\mathrm{ovs}}=5$.}
    \label{fig:proposed_control_loop_normal}
    \vspace{-1em}
\end{figure}

The proposed control structure is developed based on the multirate control scheme introduced in \cite{Tian2020MultirateHarmonic}.
It consists of a base-rate control task that operates like a DSDU control structure,
and a high-rate task that runs at an integer multiple of the base rate. Specifically,
the base-rate task is executed on the main CPU with a period of $T_{\mathrm{cpu}}$, and 
the high-rate task is executed on the auxiliary CPU, which is commonly available in modern DSPs\cite{Yu2017ImplementationofMultiSampling},
with a period of $T_{\mathrm{ovs}}=T_{\mathrm{cpu}}/K_{\mathrm{ovs}}$,
where $K_{\mathrm{ovs}} \in \mathbb{N} $ is the oversampling ratio.
Executing these tasks on separate CPUs ensures 
that the computational load of the high-rate task 
does not impact the base-rate task performance.
In addition, this approach requires minimal modifications to the existing base-rate task on the main CPU.

\figurename{\ref{fig:proposed_control_loop}} illustrates the control structure in detail.
The inverter output current, $\mathbf{i}_{s}=[i_{as}\ i_{bs}\ i_{cs}]^\top$,
and 
the point-of-common coupling (PCC) voltage, 
$\mathbf{v}_{\mathrm{pcc}}=[v_{a,\mathrm{pcc}}\ v_{b,\mathrm{pcc}}\ v_{c,\mathrm{pcc}}]^\top$,
are firstly sampled in the high-rate task.
The sampled values are delivered to the base-rate task only at the beginning of each base-rate period.
Therefore, $\mathbf{i}_{\alpha\beta,s}$ and $\mathbf{v}_{\alpha\beta,\mathrm{pcc}}$ in the base-rate task
are represented as a ZOH of those in the high-rate task.
In this paper, 
boldface variables, $\mathbf{v}$ and $\mathbf{i}$,
denote vectored voltage and current quantities, respectively.
The subscripts `$\alpha\beta$' and `$dq$' indicate quantities in stationary $\alpha\beta$ frame and synchronous $dq$ frame,
whereas those without frame subscripts represent three-phase quantities.
The superscript `*' indicates a reference value.

The base-rate task executes current control 
and grid synchronization employing
a phase-locked loop (PLL) and a dual second-order generalized integrator (DSOGI).
Meanwhile,
the high-rate task executes following functions:
\begin{enumerate}
    \item Feedforward of the $\mathbf{v}_{\alpha\beta,\mathrm{pcc}}$ to $\mathbf{v}_{\alpha\beta,s}^*$, 
    which is the voltage reference from the current controller.
    \item Computation of output pole voltage, $\mathbf{v}_n^*$, for space-vector PWM based on offset voltage injection \cite{Kim1996ANovelVoltage,Chung1998UnifiedVoltage}.
    \item Computation of grid-voltage disturbance, $\Delta \bar{\mathbf{v}}_{\alpha\beta,\mathrm{pcc}}$.
    \item Estimation of current change, $\Delta\bar{i}_{\mathrm{mag}}$, caused by the grid-voltage disturbance.
    \item Generation of reset signals, `RST', to shift PWM carrier for rapid response to grid faults.
\end{enumerate}

In normal operation without grid faults,
the operation 1) and 2) are only effective in each high-rate task execution
while operations 3) to 5) do not affect the control.
As shown in 
\figurename{\ref{fig:proposed_control_loop_normal}}, 
while the compare values in shadow registers are updated every $T_{\mathrm{ovs}}$ in the high-rate task,
the active compare values are loaded from shadow registers only at the peaks and valleys of the carrier counter.
It means that only the compare values updated at the $K_{\mathrm{ovs}}$-th high-rate task execution
within the $k$-th base-rate period
are applied to the PWM module to generate gating signals during the entire $(k+1)$-th base-rate period.
Consequently, the inverter operates similarly to the DSDU scheme from the current-control perspective.

\subsection{Estimation of Grid-Voltage Disturbance} \label{sec:computation_of_grid_voltage_disturbance}
This section describes the first sub-block in the high-rate task and the oversampling window counter, 
as shown in \figurename{\ref{fig:proposed_control_loop}}, 
which estimates the grid-voltage disturbance.

Under the assumption that the inverter accurately 
synthesize the output voltage reference, $\mathbf{v}_{n}^*=[v_{an}^*\  v_{bn}^*\ v_{cn}^*]^\top$, over 
a base-rate period,
the inverter output current dynamics can be represented in $\alpha\beta$ frame as follows:
\begin{equation}
    \frac{d}{dt}\mathbf{i}_{\alpha\beta,s} = \frac{1}{L_s}\left( \mathbf{v}_{\alpha\beta,n}^* - \mathbf{v}_{\alpha\beta,\mathrm{pcc}} - R_s \mathbf{i}_{\alpha\beta,s} \right),
    \label{eq:current_dynamics}
\end{equation}
where $L_s$ and $R_s$ are the inductance and resistance of the filter impedance, $Z_f$, respectively.
$\mathbf{v}_{\alpha\beta,n}^*$
is the $\alpha\beta$-frame representation of $\mathbf{v}_{n}^*$ 
and can be expressed as follows:
\begin{equation}
    \mathbf{v}_{\alpha\beta,n}^* = \mathbf{v}_{\alpha\beta,s}^* + \mathbf{v}_{\alpha\beta,\mathrm{pcc}}^*,
    \label{eq:output_voltage_reference}
\end{equation}
where $\mathbf{v}_{\alpha\beta,\mathrm{pcc}}^*$ is the feedforward PCC voltage reference.

From the perspective of the switching-period average,
the PCC voltage is effectively cancelled out by the feedforward term during normal operation,
resulting in current dynamics primarily governed by the $\mathbf{v}_{\alpha\beta,s}^*$.
This implies that the following approximation holds:
\begin{align}
    \Delta{\bar{\mathbf{v}}}_{\alpha\beta,\mathrm{pcc}} \Big|_{k+1}^{k+2}&=
    \bar{\mathbf{v}}_{\alpha\beta,\mathrm{pcc}}^*[k,K_{\mathrm{ovs}}] \nonumber \\
    &\quad-\frac{1}{T_{\mathrm{cpu}}}\int_{(k+1)T_{\mathrm{cpu}}}^{(k+2)T_{\mathrm{cpu}}}\mathbf{v}_{\alpha\beta,\mathrm{pcc}}(t) dt \nonumber \\
    &\approx 0,
    \label{eq:grid_disturbance_nofault}
\end{align}
where  $\bar{\mathbf{v}}_{\alpha\beta,\mathrm{pcc}}^*[k,l]$
represent the feedforward PCC voltage reference 
calculated at the $l$-th high-rate task execution
within the $k$-th base-rate period.
$\Delta{\bar{\mathbf{v}}}_{\alpha\beta,\mathrm{pcc}} \big|_{k+1}^{k+2}$
denotes the voltage difference averaged from the $(k+1)$-th to the $(k+2)$-th base-rate task instants.
Since $\bar{\mathbf{v}}_{\alpha\beta,\mathrm{pcc}}^*[k,K_{\mathrm{ovs}}]$
is the measured PCC voltage right before the $(k+1)$-th base-rate period,
$\Delta{\bar{\mathbf{v}}}_{\alpha\beta,\mathrm{pcc}} \big|_{k+1}^{k+2}$
is nearly zero during normal operation without grid faults.

However, when a grid fault occurs,
the approximation in \eqref{eq:grid_disturbance_nofault} no longer holds.
It induces a significant $\Delta{\bar{\mathbf{v}}}_{\alpha\beta,\mathrm{pcc}} \big|_{k+1}^{k+2}$
and eventually leads to transient overcurrent.
To detect such grid voltage disturbances,
the oversampling window counter and the first sub-block are implemented in the high-rate task as shown in \figurename{\ref{fig:proposed_control_loop}}.

The value of the oversampling window counter, $\mathrm{CTR}_{\mathrm{ovs},w}$,
increments by one at each high-rate task execution indicating the sequence number of the present high-rate task execution within a single base-rate period.
Once it reaches $K_{\mathrm{ovs}}$,
it generates a pulse signal, $U_{\mathrm{end}}$, indicating the last high-rate task execution within the present base-rate period,
and resets to one at the next high-rate task execution.

When $U_{\mathrm{end}}$ is asserted,
the update flag in the first sub-block, $U_{\mathrm{ff}}$, is set to one.
Consequently, the feedforward memory updates and 
stores the value of $\bar{\mathbf{v}}_{\alpha\beta,\mathrm{pcc}}^*[k,K_{\mathrm{ovs}}]$,
which corresponds to the PCC voltage reference synthesized by the inverter during the $(k+1)$-th base-rate period.
By comparing this stored value with the PCC voltage sampled in the subsequent high-rate task executions,
the instantaneous grid-voltage disturbance within a base-rate period is estimated as follows:
\begin{equation}
    \Delta\hat{\mathbf{v}}_{\alpha\beta,\mathrm{pcc}}[k+1,l] 
    = \bar{\mathbf{v}}_{\alpha\beta,\mathrm{pcc}}^*[k,K_{\mathrm{ovs}}] - \mathbf{v}_{\alpha\beta,\mathrm{pcc}}[k+1,l] ,
    \label{eq:estimated_grid_voltage_disturbance}
\end{equation}
where $ \Delta\hat{\mathbf{v}}_{\alpha\beta,\mathrm{pcc}}[k+1,l] $ 
denotes the estimated grid-voltage disturbance at the $l$-th high-rate task execution within the $(k+1)$-th base-rate period.
$l$ is in the range of $1 \leq l \leq K_{\mathrm{ovs}}$.
After the computation of \eqref{eq:estimated_grid_voltage_disturbance},
$\Delta\hat{\mathbf{v}}_{\alpha\beta,\mathrm{pcc}}$ is delivered to the second sub-block 
to estimate transient current change.

\subsection{Estimation of Current Change within a Switching Period} \label{sec:estimation_of_transient_current_change_within_a_switching_period}
This section details the second sub-block in the high-rate task, as shown in \figurename{\ref{fig:proposed_control_loop}}, 
which estimates the transient current change caused by the detected grid voltage disturbance.

At every high-rate task execution within the $(k+1)$-th base-rate period,
the remaining time of the present base-rate period, $T_{\mathrm{rem}}$,
can be defined as follows:
\begin{equation}
    T_{\mathrm{rem}} = \left(1 - \frac{l}{K_{\mathrm{ovs}}}\right)T_{\mathrm{cpu}}.
    \label{eq:remaining_time_current_base_rate}
\end{equation}
where $l$ is the sequence number of the present high-rate task execution within the $(k+1)$-th base-rate period,
which is implemented by the oversampling window counter, $\mathrm{CTR}_{\mathrm{ovs},w}$.

Assuming the filter resistance is negligible during the short transient period,
the inverter output current dynamics in \eqref{eq:current_dynamics} can be approximated as follows:
\begin{align}
    \frac{d}{dt}\mathbf{i}_{\alpha\beta,s} &\approx \frac{1}{L_s}\left( \mathbf{v}_{\alpha\beta,n}^* - \mathbf{v}_{\alpha\beta,\mathrm{pcc}} \right).
    \label{eq:current_dynamics_approx}
\end{align}

By integrating \eqref{eq:current_dynamics_approx}
over $T_{\mathrm{rem}}$
from the $l$-th high-rate task execution to the end of the $(k+1)$-th base-rate period,
the total current change can be derived. However,
this direct integration includes both the current variation driven by the current controller and the PCC voltage disturbance.
Therefore, it is necessary to isolate the contribution of the PCC voltage disturbance.
From this perspective,
it is worth noting that $\Delta\hat{\mathbf{v}}_{\alpha\beta,\mathrm{pcc}}[k+1,l]$ in \eqref{eq:estimated_grid_voltage_disturbance} 
effectively approximates 
the instantaneous grid-voltage disturbance within a base-rate period
and isolates the fault-induced current change.
This is because $ \bar{\mathbf{v}}_{\alpha\beta,\mathrm{pcc}}^*[k,K_{\mathrm{ovs}}]$ 
serves not only as the latched feedforward voltage reference but also as a representation of the pre-fault PCC voltage.

To isolate the fault-induced current variation,
$\Delta\hat{\mathbf{v}}_{\alpha\beta,\mathrm{pcc}}[k+1,l]$ 
is utilized as follows:
\begin{equation}
    \Delta\hat{\mathbf{i}}_{\alpha\beta,s}[k+1,l]
    = \frac{T_{\mathrm{rem}}}{L_s}\Delta\hat{\mathbf{v}}_{\alpha\beta,\mathrm{pcc}}[k+1,l]
    \label{eq:estimated_current_change_due_to_grid_disturbance}
\end{equation}
where $\Delta\hat{\mathbf{i}}_{\alpha\beta,s}[k+1,l]$
represents the estimated current change caused by the grid-voltage disturbance
for the remaining time $T_{\mathrm{rem}}$
at the $l$-th high-rate task execution within the $(k+1)$-th base-rate period.
In \eqref{eq:estimated_current_change_due_to_grid_disturbance},
it is assumed that $\Delta\hat{\mathbf{v}}_{\alpha\beta,\mathrm{pcc}}[k+1,l]$ 
remains constant during the remaining time in this paper to simplify the estimation and reduce the computational burden.
Although the disturbance may vary continuously during the remaining time, 
this assumption is justified as the reduced computational burden allows for a higher oversampling ratio,
thereby enabling a rapid fault response.

Once $\Delta\hat{\mathbf{i}}_{\alpha\beta,s}[k+1,l]$ in \eqref{eq:estimated_current_change_due_to_grid_disturbance} is obtained,
its magnitude, $\Delta\hat{i}_{\mathrm{mag}}$, is calculated as L2-norm and delivered to the third sub-block for PWM carrier shift decision.

\subsection{PWM Carrier Shift for Rapid Response to Grid Faults}\label{sec:pwm_carrier_shift}

\begin{figure}[!t]
    \centering
    \includegraphics[width=1.0\linewidth]{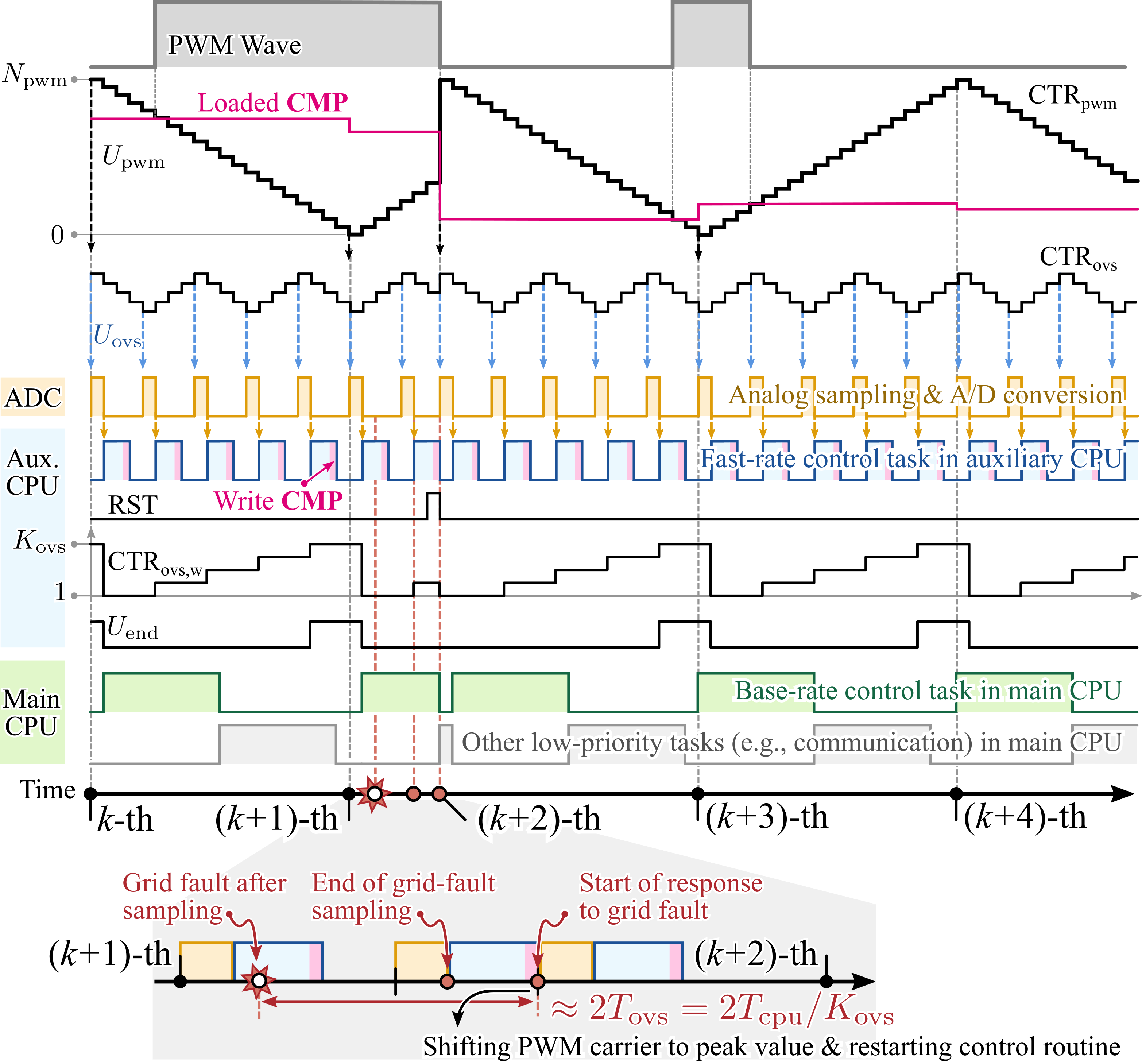}
    \caption{PWM update timing diagram of the proposed control structure during grid-fault condition when $K_{\mathrm{ovs}}=5$.}
    \label{fig:proposed_control_loop_worstcase}
    \vspace{-1em}
\end{figure}

This section details the third sub-block in the high-rate task, as shown in \figurename{\ref{fig:proposed_control_loop}}, 
which determines the PWM carrier shift and control period reset in response to grid faults.

In the third sub-block,
there is a predefined current-change threshold, $\Delta I_{\mathrm{mag}}$.
When the estimated current change magnitude, $\Delta\hat{i}_{\mathrm{mag}}$,
exceeds this threshold,
it represents a significant grid-voltage disturbance that may lead to transient overcurrent.
In this case,
to shift the PWM carrier and reset the control period at the next high-rate task execution,
a reset signal, `RST', is set to 1 as follows:
\begin{equation}
    \mathrm{RST}=\begin{cases}
        1, & \text{if } \Delta\hat{i}_{\mathrm{mag}}=\|\Delta\hat{\mathbf{i}}_{\alpha\beta,s}[k+1,l]\|_2 > \Delta I_{\mathrm{mag}} \\
        0, & \text{otherwise}.
    \end{cases}
    \label{eq:RST_asserted}
\end{equation}

When `RST' is asserted,
the voltage feedforward memory in the high-rate task is updated immediately, 
because the update flag, $U_{\mathrm{ff}}$,
is the result of the logical `OR' operation between $U_{\mathrm{end}}$ and `RST'.
Consequently, the PCC voltage feedforward memory,
$\bar{\mathbf{v}}_{\alpha\beta,\mathrm{pcc}}^*[k+1,K_{\mathrm{ovs}}]$,
is updated either at the end of the base-rate period or when a grid fault is detected.
Moreover, before the end of the current high-rate task execution,
$\mathbf{CMP}$ values in the shadow registers of the PWM module are updated
based on the latest sampled $\mathbf{v}_{\alpha\beta,\mathrm{pcc}}$ and computed $\mathbf{v}_{\alpha\beta,s}$.
After writing the updated $\mathbf{CMP}$ values to the shadow registers,
$\mathrm{CTR}_{\mathrm{ovs},w}$ is set to 1 so that the next high-rate task execution
corresponds to the first high-rate task execution within the new base-rate period.

As the last step of the third sub-block when the `RST' signal is asserted,
`RST' signal triggers simultaneous resets of both the PWM carrier counter and the oversampling interrupt counter to their peak values.
This action causes the PWM carrier counter, $\mathrm{CTR}_{\mathrm{pwm}}$,
and the oversampling interrupt counter, $\mathrm{CTR}_{\mathrm{ovs}}$,
to start counting down from their peak values, $N_{\mathrm{pwm}}$ and $N_{\mathrm{ovs}}$, respectively,
resulting in an immediate load of the active compare values from the shadow registers.
It leads to the prompt reaction of the inverter output voltage to the grid fault within sub-switching period,
thereby effectively mitigating transient overcurrent.

In \figurename{\ref{fig:proposed_control_loop_worstcase}},
the worst-case grid-fault scenario when the proposed control structure is adopted is illustrated.
Here, it is assumed that a grid fault occurs just after $(k+1)$-th base-rate period begins,
when  the PWM compare values are loaded, and the grid-side voltage and current are sampled.
Since the fault occurs right after the analog-to-digital conversion of the first high-rate task execution within this base-rate period,
the digital controller detects the fault at the second high-rate task execution when $\mathrm{CTR}_{\mathrm{ovs},w}=2$.

At this moment, the compare values in the shadow registers 
are updated based on the newly sampled grid voltage and latest voltage reference from the current controller.
Through the computation of \eqref{eq:estimated_grid_voltage_disturbance},\eqref{eq:estimated_current_change_due_to_grid_disturbance},
and \eqref{eq:RST_asserted},
`RST' is asserted because the estimated current change, $\Delta\hat{i}_{\mathrm{mag}}$,
exceeds the predefined threshold, $\Delta I_{\mathrm{mag}}$.
Therefore, the voltage feedforward memory in the first sub-block is updated immediately,
and the oversampling window counter, $\mathrm{CTR}_{\mathrm{ovs},w}$,
is reset to 1 before the end of the second high-rate task execution.
Finally, `RST' triggers the resets of both the PWM carrier counter and the oversampling interrupt counter
to immediately load the active compare values from the shadow registers and restart the control routines.
As a result, the inverter is able to respond to the grid fault within the two times of $T_{\mathrm{ovs}}$.
By increasing the oversampling ratio, $K_{\mathrm{ovs}}=T_{\mathrm{cpu}}/T_{\mathrm{ovs}}$,
the response time can be further reduced,
enabling rapid suppression of transient overcurrent during grid faults.
\section{Experimental Results} \label{sec:experimental_results}
This section presents the experimental results of the proposed control structure under various grid-fault scenarios.
First, the experimental setup is described to provide the implementation details of the proposed control structure.
Then the effectiveness of the proposed control structure is validated through experiments under symmetric voltage sags, asymmetric voltage sags, and phase jumps.

\subsection{Experimental Setup} \label{sec:experimental_setup}
\begin{figure}[!t]
    \centering
    \includegraphics[width=0.95\linewidth]{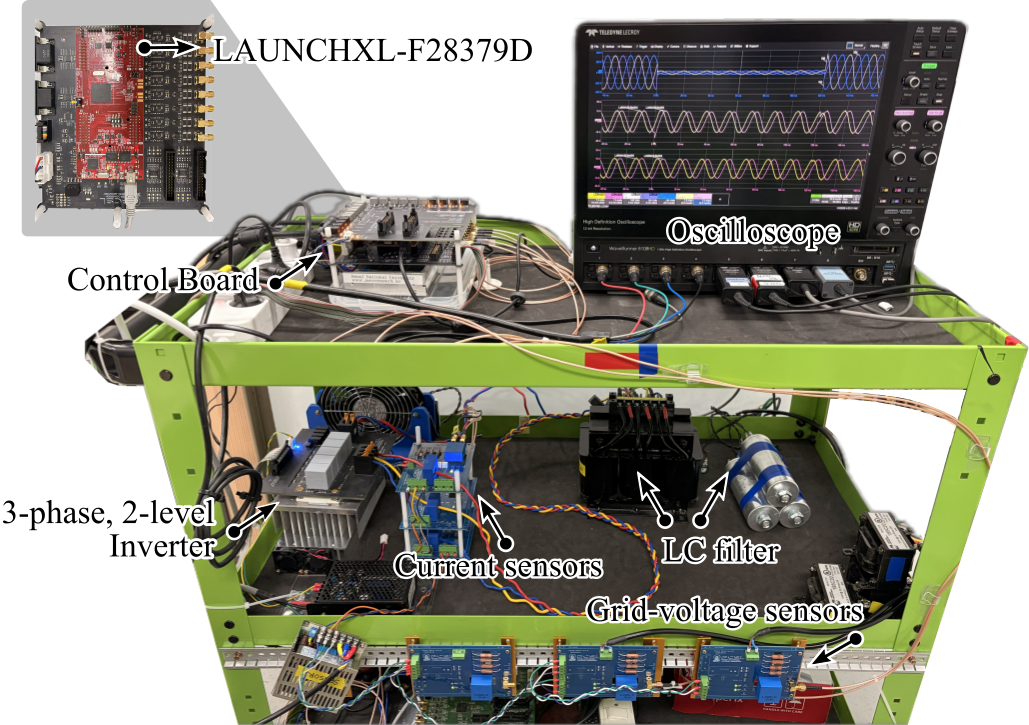}
    \caption{Experimental setup.}
    \label{fig:experimental_setup}
    \vspace{0em}
\end{figure}

\begin{table}[t]
        \begin{threeparttable}
            \caption{\textsc{System Parameters for Experiment}}
            \label{tab:params_sys}
            \centering
            {\renewcommand{\arraystretch}{1.1}
        
            \begin{tabular*}{\columnwidth}{@{\extracolsep{\fill}}l@{\hspace{0.2em}}l@{\hspace{0.5em}}l@{\hspace{0.2em}}l}
                \toprule
                \multicolumn{4}{c}{AC Grid} \\ 
                \midrule
                Nominal grid frequency & 60\,Hz   & Grid line-to-line voltage\tnote{*}     & 220\,V\textsubscript{rms}\\
                \midrule\midrule
                \multicolumn{4}{c}{Grid-tied Inverter} \\ 
                \midrule
                Rated power & 4\,kW & Rated line current & 15\,A\\
                DC-link voltage & 400\,V & Filter inductance, $L_f$\tnote{**} & \makecell[l]{3.4\,mH\,(0.1\,p.u.)}\\
                Filter resistance, $R_f$ & 12.5\,m$\Omega$ & Filter capacitance, $C_f$\tnote{**} & \makecell[l]{55\,\si{\micro\farad}\,(3.9\,p.u.)}\\
                Switching frequency & 3.5\,kHz & Base-rate frequency & 7.0\,kHz \\ 
                High-rate frequency & 105\,kHz & Oversampling ratio, $K_{\mathrm{ovs}}$ & 15 \\
                \bottomrule
            \end{tabular*}%
        
            \scriptsize
            \begin{tablenotes}
            \item[*]{The subscript `rms' denotes root-mean-square value at the rated operation.}
            \item[**]{The p.u. values are per-unit impedance based on the base impedance of the inverter.}
            \end{tablenotes}
            \vspace{0em}
        }
        \end{threeparttable}
        \vspace{-1em}
\end{table}

\figurename{\ref{fig:experimental_setup}} shows the experimental setup used 
to validate the proposed current-limiting strategy.
A 4-kW grid-tied inverter is connected to a programmable grid simulator,
MX-30 from California Instruments\textsuperscript{\copyright}
through an LC filter.
The proposed control structure is implemented on a LAUNCHXL-F28379D board from Texas Instruments\textsuperscript{\copyright},
which is a low-cost development kit based on the TMS320F28379D DSP\cite{launchxlF28379d}.
The high-rate task is implemented on the control-law accelerator (CLA) co-processor in the DSP\cite{instruments2024tms320f2837xd},
while the base-rate task runs on the main CPU.
Detailed system parameters are summarized in \tablename{\ref{tab:params_sys}}.

The objective of the experiments is to demonstrate that 
the proposed control structure can effectively limit transient overcurrents during sub-switching period.
Therefore, in each experiment, the inverter is operated 
to inject rated current under normal conditions before the grid fault occurs.
During the fault-ride-through (FRT) operation, 
the current reference is kept constant.
Note that the current reference profile during FRT 
can be determined by the outer control loop according to grid codes and system requirements,
which is beyond the scope of this paper.

The effectiveness of the proposed control structure is 
validated by comparing the transient overcurrent with that of the conventional DSDU 
control structure under the same conditions.
The p.u. current divisions in the experimental results 
are based on the rated current of 15\,A (1\,p.u.).

\begin{figure}[t]
    \centering
    \includegraphics[width=1.0\linewidth]{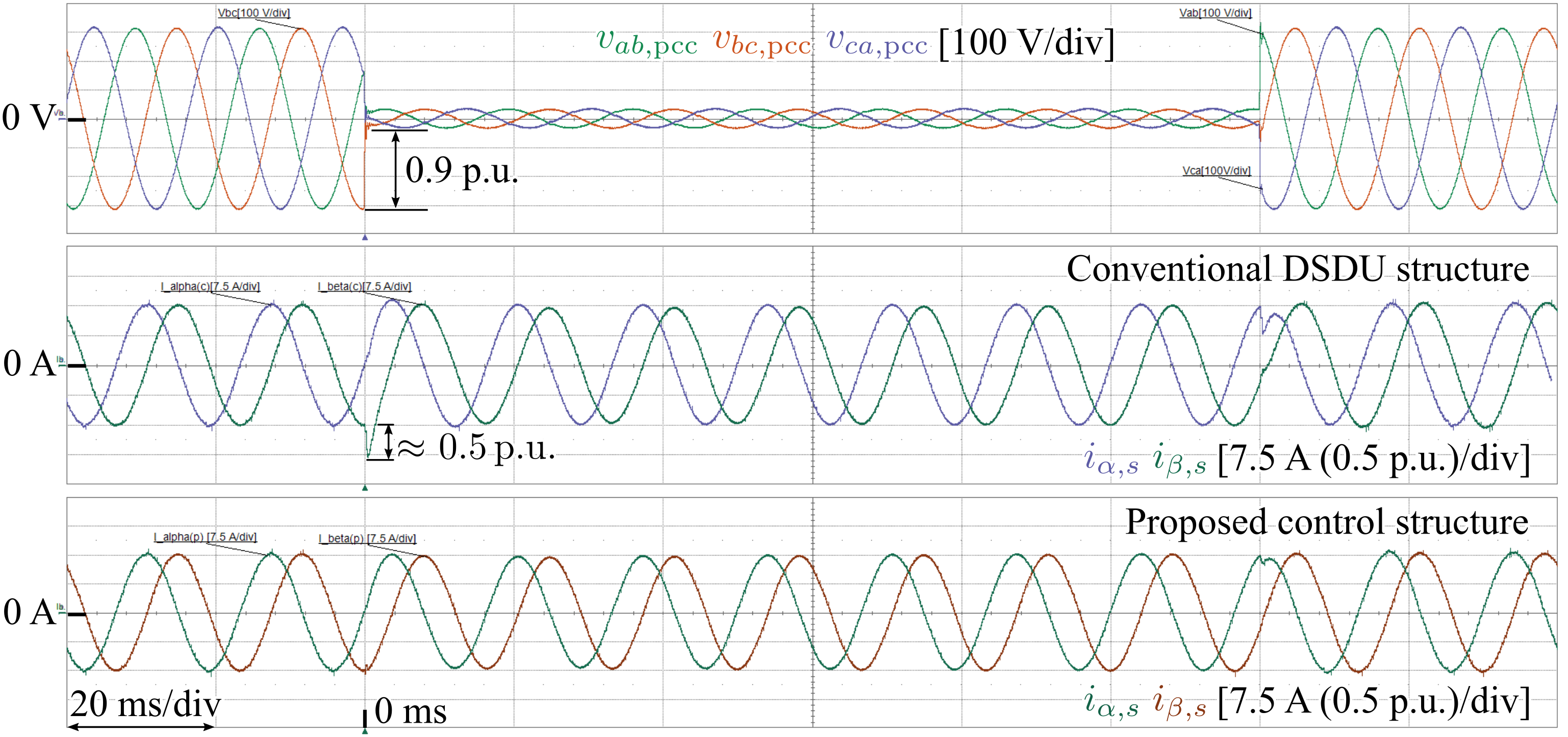}
    \caption{Experimental results under a symmetric voltage sag of 90\% depth.}
    \label{fig:exp_sym_voltage_sag}
    \vspace{0em}
\end{figure}

\begin{figure}[t!]
    \vspace{-0.0em}  
    \centering
    \begin{subfigure}[b]{1.0\linewidth}
        \includegraphics[width=1.0\linewidth]{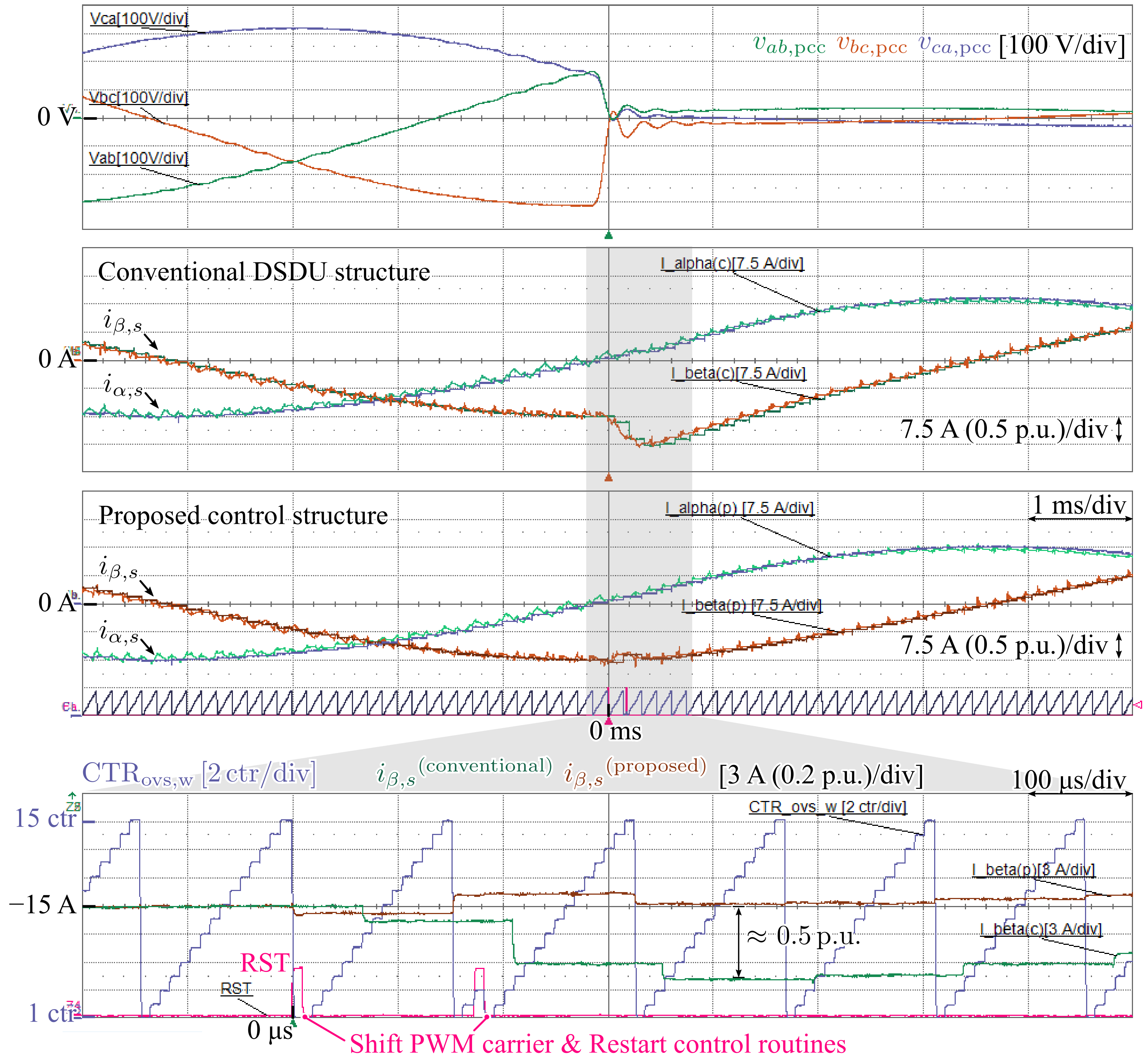}
        \caption{}
        \label{fig:exp_sym_voltage_sag_start_zoom}
        \vspace{1em}
    \end{subfigure}
    \begin{subfigure}[b]{1.0\linewidth}
        \includegraphics[width=1.0\linewidth]{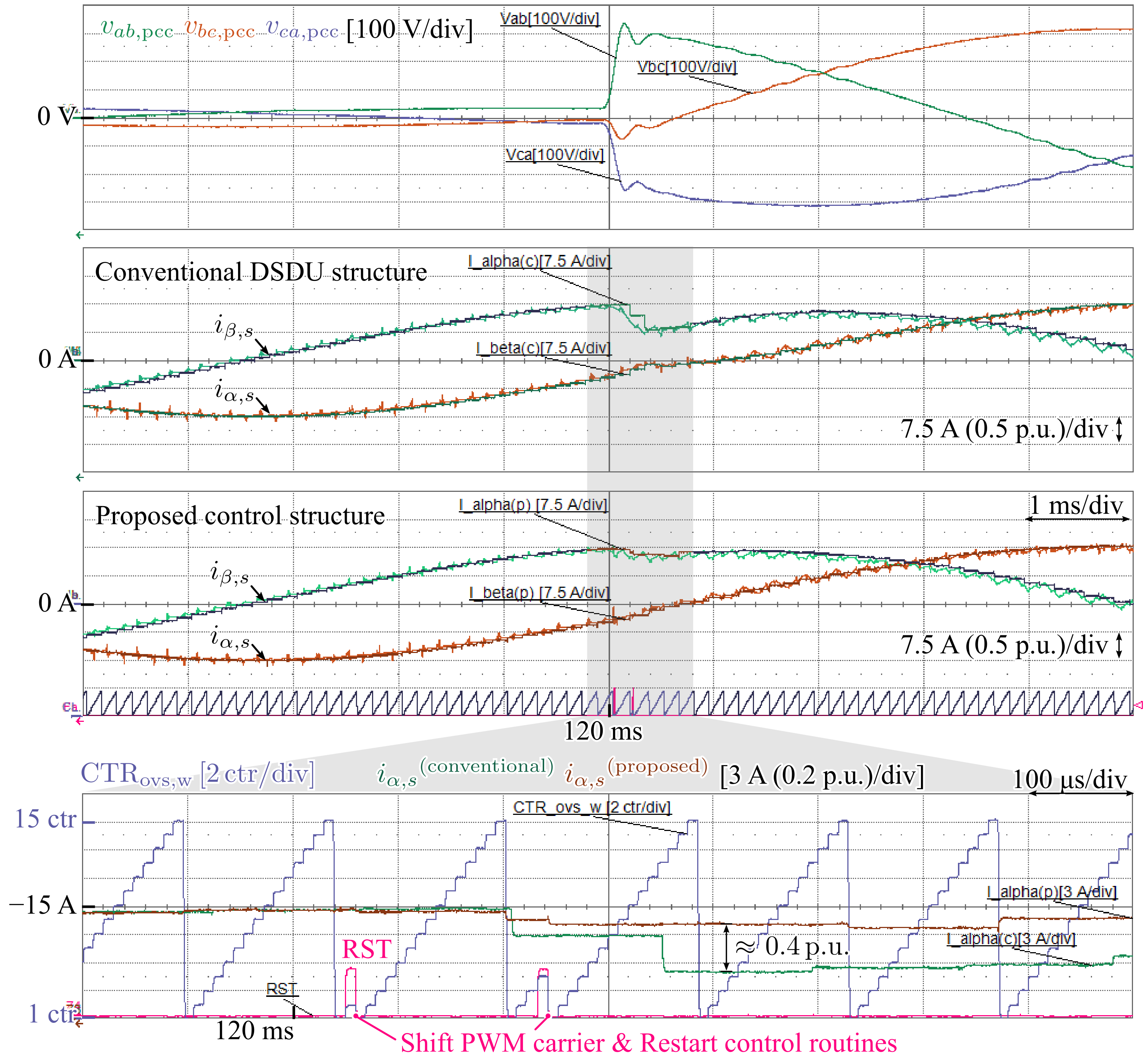}
        \caption{}
        \label{fig:exp_sym_voltage_sag_end_zoom}
    \end{subfigure}
    \caption{Zoomed waveform in \figurename{\ref{fig:exp_sym_voltage_sag}}: (a) at the start of the voltage sag and (b) at the end of the voltage sag.}
    \label{fig:exp_sym_voltage_sag_zoom}
   \vspace{-1.0em}  
\end{figure}

\subsection{Symmetric Voltage Sag} \label{sec:symm_voltage_sag}

\figurename{\ref{fig:exp_sym_voltage_sag}}
shows the experimental results under a symmetric voltage sag of 90\% depth.
At $t=0$\,s,
the grid voltage suddenly drops to 10\% of its nominal value.
After that, it recovers to the nominal value at $t=0.12$\,s.
The first row of \figurename{\ref{fig:exp_sym_voltage_sag}}
shows the line-to-line voltage during the voltage sag.
The second row of \figurename{\ref{fig:exp_sym_voltage_sag}} 
shows the inverter output current response at 
the stationary $\alpha\beta$ frame
under the conventional DSDU control structure,
while the third row shows that under the proposed control structure.
Although 
both control structures successfully limit the steady-state current
to the rated value during the voltage sag,
the transient overcurrent under the conventional DSDU control structure
reaches up to 1.5 p.u. which may trigger the overcurrent protection.
On the other hand,
the proposed control structure effectively limits the transient overcurrent.

\figurename{\ref{fig:exp_sym_voltage_sag_zoom}}
shows the zoomed waveform at the start and the end of the voltage sag.
In \figurename{\ref{fig:exp_sym_voltage_sag_start_zoom}},
it is observed that 
the transient overcurrent
is uncontrolled during the consecutive two base-rate periods
right after the voltage sag occurs
under the conventional DSDU control structure.
This is because 
the inherent delay of the conventional control structure
prevents the immediate response to the sudden voltage drop,
as mentioned in Section \ref{sec:introduction}.
On the other hand,
the proposed control structure
effectively responds to the voltage sag
much faster than a single base-rate period
by shifting the PWM carrier.
This is the main difference 
from the control-based
methods in \cite{Dong2023APLL_lessVoltage, Kawashima2024UltraRobust1MHz},
which modify the duty cycle without changing the carrier position,
even when the deadbeat control or multisampling techniques are adopted.

In the last row of \figurename{\ref{fig:exp_sym_voltage_sag_start_zoom}},
the detailed operation of the proposed control structure is illustrated.
The $i_{\beta}$ component of the inverter output current,
which exhibits the largest transient deviation due to the voltage sag,
is plotted together with the 
oversampling window counter, CTR$_{\mathrm{ovs,w}}$, and the reset signal, `RST'.
Once the grid fault is detected by the high-rate task,
the CLA immediately triggers the `RST'
signal to shift the PWM carrier and restart the control routines.
As a result, CTR$_{\mathrm{ovs,w}}$
resets to 1,
and the duty cycle is automatically loaded 
based on the latest grid voltage measurement.
This process enables the rapid response to grid faults.

Similarly, in \figurename{\ref{fig:exp_sym_voltage_sag_end_zoom}},
at the end of the voltage sag,
the proposed control structure quickly restores the normal operation
by shifting the PWM carrier again.
As a result,
the transient overcurrent at the recovery instant
is also effectively limited,
while the conventional DSDU control structure
exhibits a significant current dip due to the delayed response.
\begin{figure}[t]
    \centering
    \includegraphics[width=1.0\linewidth]{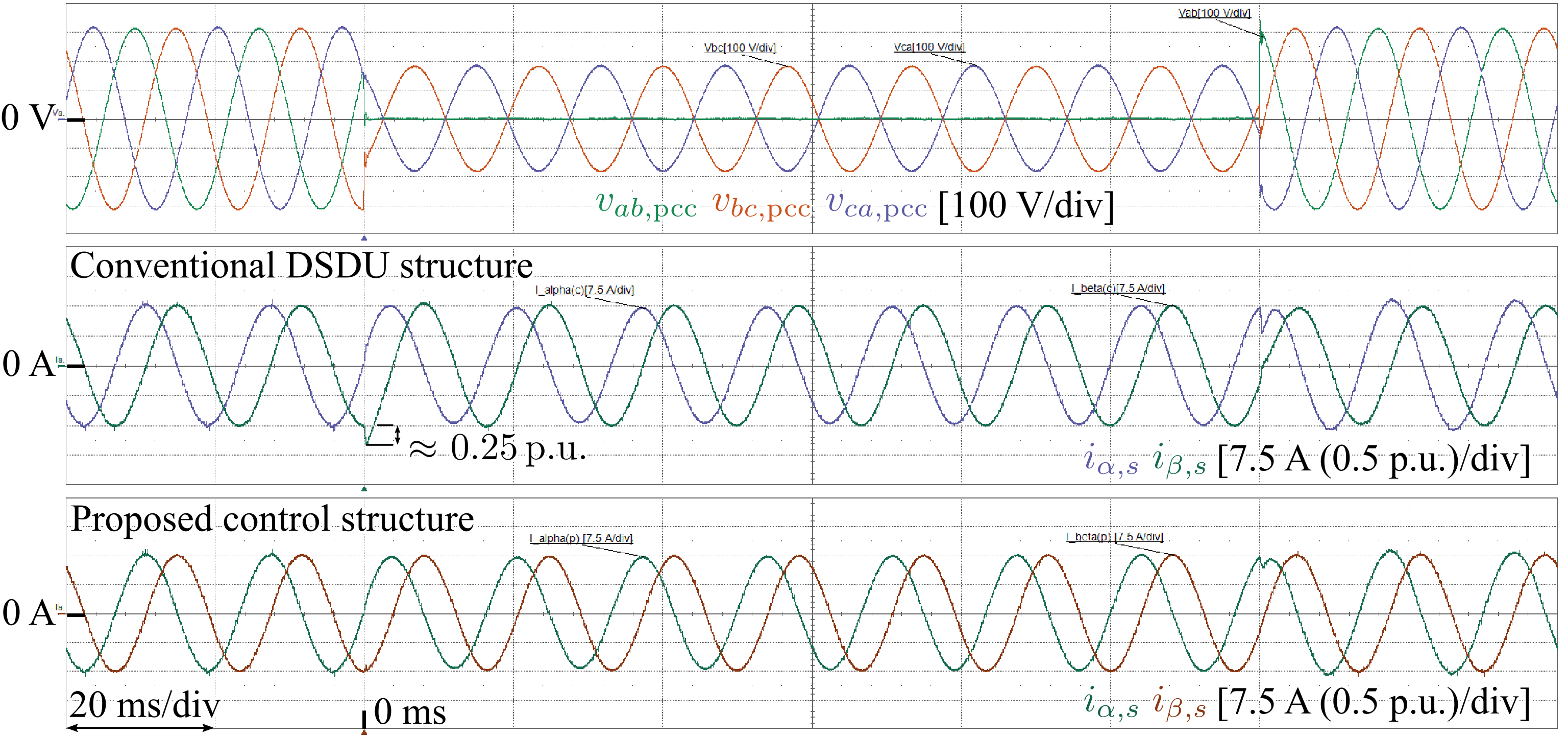}
    \caption{Experimental results under phase-A and phase-B short circuit.}
    \label{fig:exp_l2l_voltage_sag}
    \vspace{0em}
\end{figure}

\begin{figure}[t!]
    \vspace{-0.0em}  
    \centering
    \begin{subfigure}[b]{1.0\linewidth}
        \includegraphics[width=1.0\linewidth]{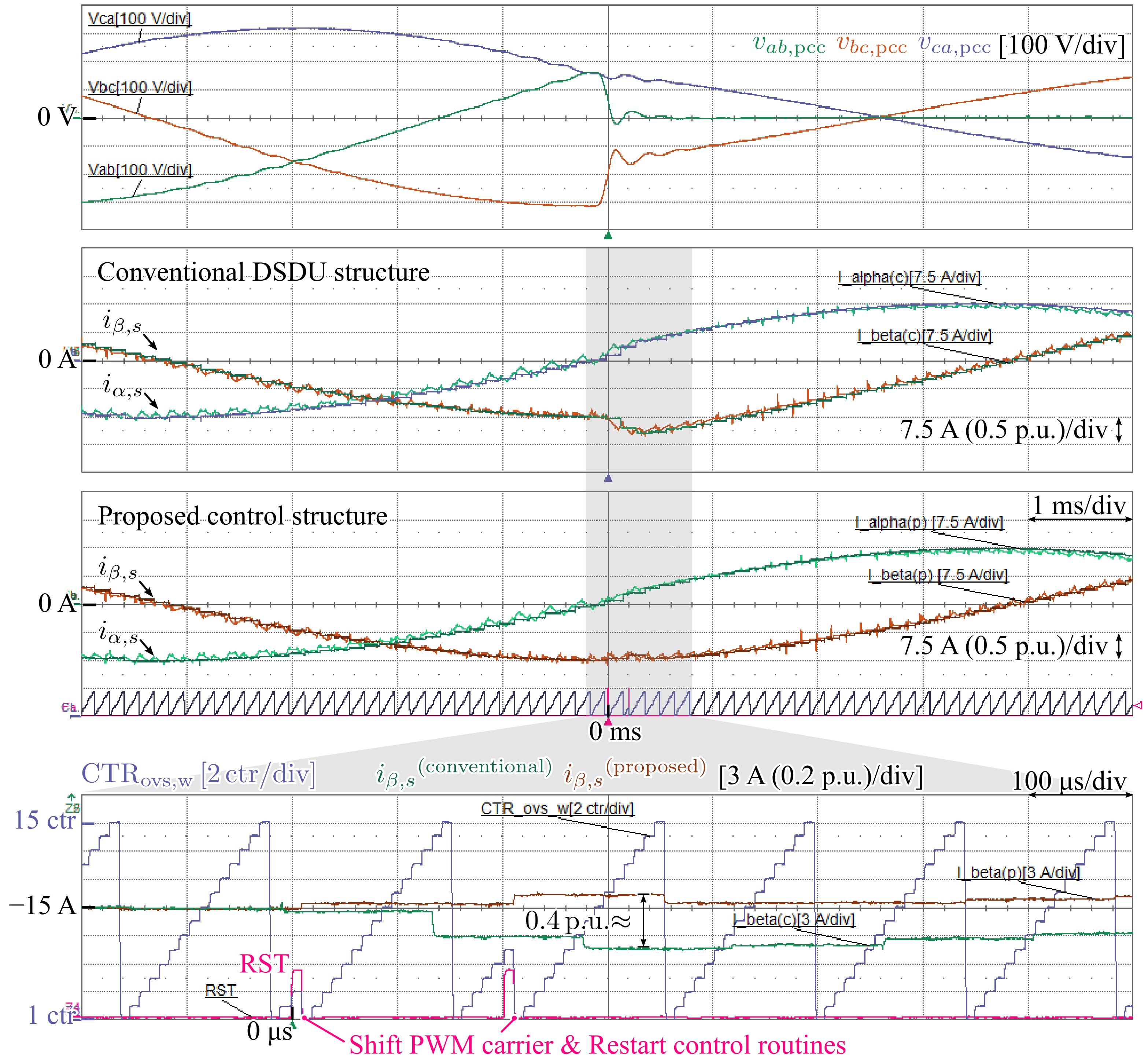}
        \caption{}
        \label{fig:exp_l2l_voltage_sag_start_zoom}
        \vspace{1em}
    \end{subfigure}
    \begin{subfigure}[b]{1.0\linewidth}
        \includegraphics[width=1.0\linewidth]{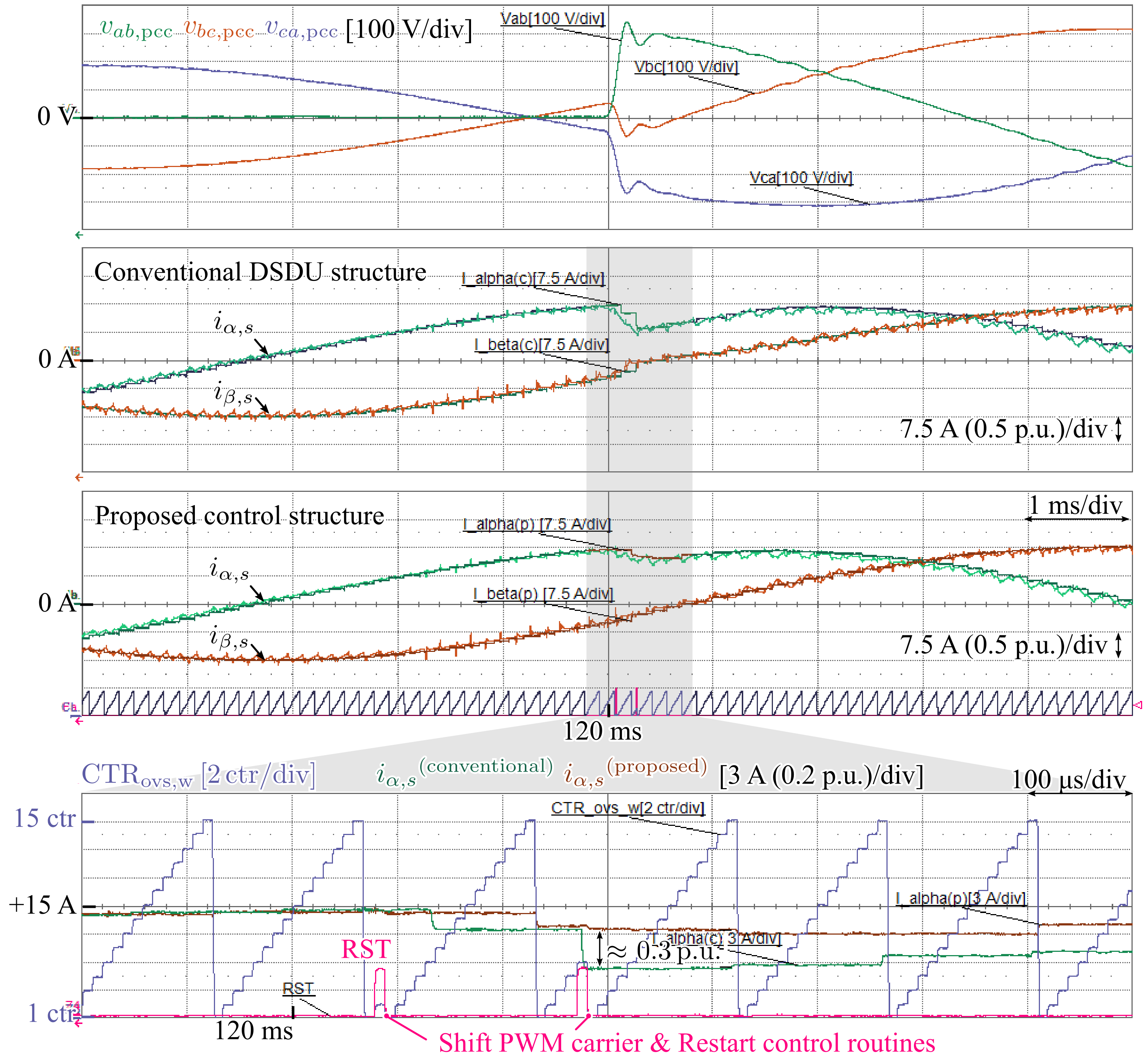}
        \caption{}
        \label{fig:exp_l2l_voltage_sag_end_zoom}
    \end{subfigure}
    \caption{Zoomed waveform in \figurename{\ref{fig:exp_l2l_voltage_sag}}: (a) at the start of the grid fault and (b) at the end of the grid fault.}
    \label{fig:exp_l2l_voltage_sag_zoom}
   \vspace{-1.0em}  
\end{figure}

\subsection{Asymmetric Voltage Sag} \label{sec:asymm_voltage_sag}

\begin{figure*}[!t]
    \vspace{-0.0em}  
    \centering
    \hfill
    \begin{subfigure}[b]{0.49\linewidth}
        \includegraphics[width=1.0\linewidth]{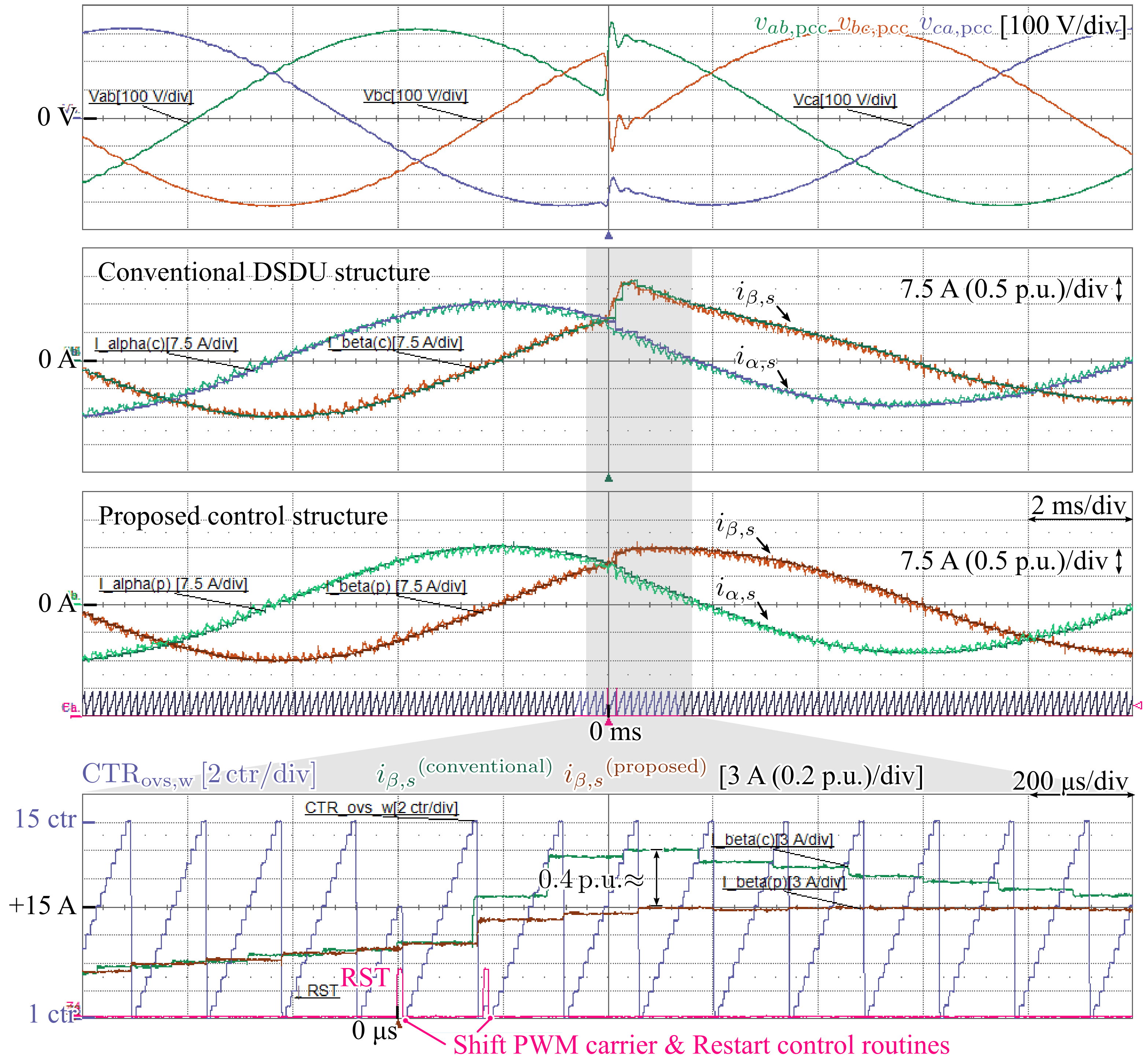}
        \caption{}
        \label{fig:exp_phase_jump_neg60}
    \end{subfigure}
    \hfill
    \begin{subfigure}[b]{0.49\linewidth}
        \includegraphics[width=1.0\linewidth]{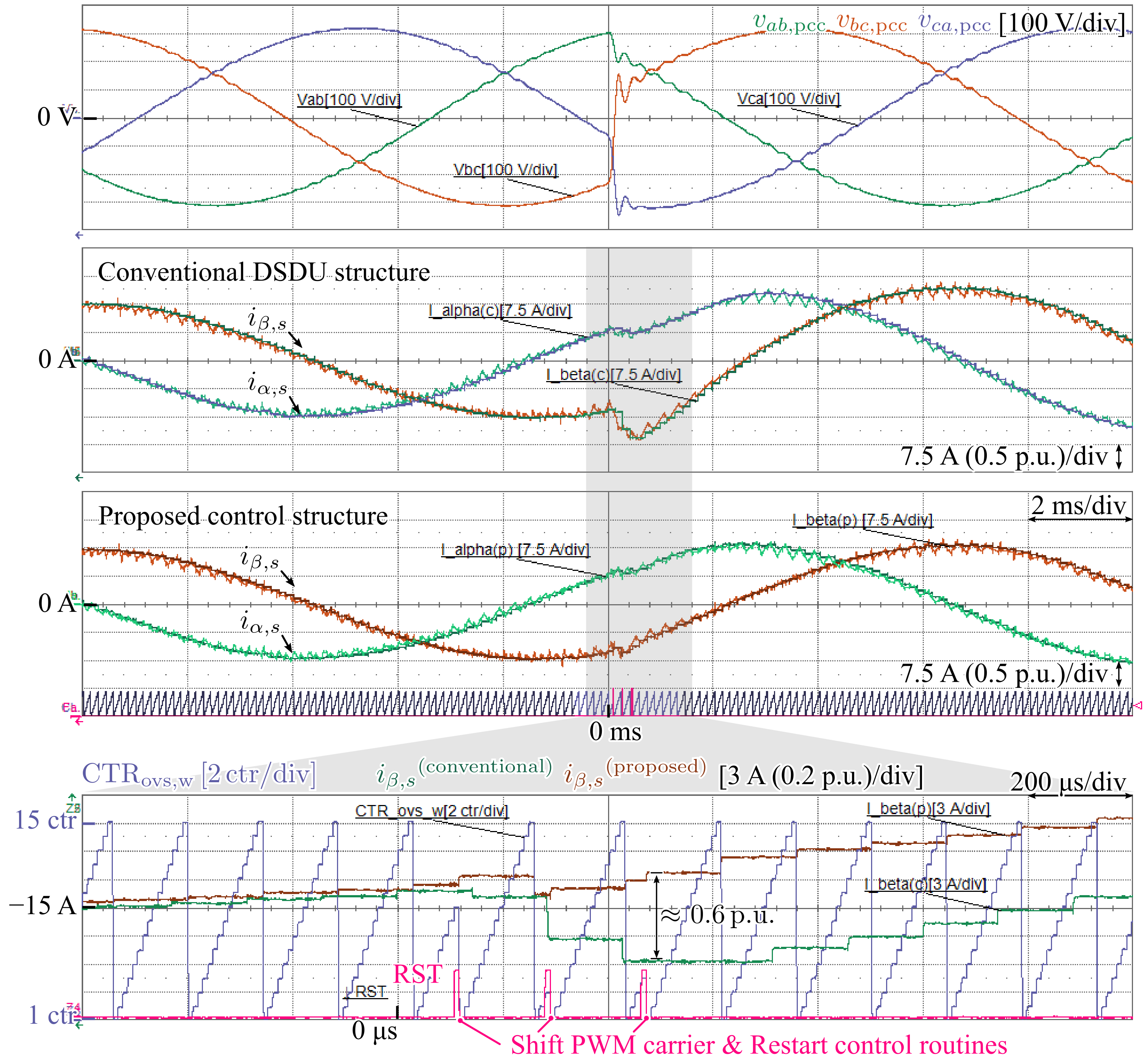}
        \caption{}
        \label{fig:exp_phase_jump_pos60}
    \end{subfigure}
    \hfill
    \caption{Experimental results under a phase jump of (a) $-$60$^\circ$ and (b) $+$60$^\circ$.}
    \label{fig:exp_phase_jump}
   \vspace{-1.0em}  
\end{figure*}

\figurename{\ref{fig:exp_l2l_voltage_sag}}
shows the 
experimental results under an asymmetric voltage sag caused by a phase-A and phase-B short circuit.
At $t=0$\,s,
the line-to-line voltage between phase A and phase B suddenly drops to nearly 0\,V.
After that, it recovers to the nominal value at $t=0.12$\,s.
The first row of \figurename{\ref{fig:exp_l2l_voltage_sag}}
shows the line-to-line voltage during the grid fault.
The second row of \figurename{\ref{fig:exp_l2l_voltage_sag}} 
shows the inverter output current response at 
the stationary $\alpha\beta$ frame
under the conventional DSDU control structure,
while the third row shows that under the proposed control structure.

Similarly to the symmetric voltage sag case,
the proposed control structure
effectively responds to the sudden voltage change,
while the conventional DSDU control structure
exhibits uncontrolled transient overcurrents
due to the inherent control delay.
\figurename{\ref{fig:exp_l2l_voltage_sag_zoom}} 
shows the zoomed waveform at the start and the end of the grid fault.
As mentioned in Section \ref{sec:symm_voltage_sag},
it is observed that
the PWM carrier shifting and control routine restarting
are triggered by the `RST' signal from the CLA
immediately after the fault detection,
enabling the rapid response to the grid fault.

\subsection{Phase Jump}\label{sec:phase_Jump}
\figurename{\ref{fig:exp_phase_jump}} shows the
experimental results 
under a phase jump grid fault.
In \figurename{\ref{fig:exp_phase_jump_neg60}},
the grid-voltage angle suddenly jumps by $-$60$^\circ$ at $t=0$\,s.
Due to the sudden angle change,
the inverter output current in the second row
exhibits a significant transient overcurrent
under the conventional DSDU control structure.
In contrast,
the proposed control structure 
reduces the transient overcurrent by 0.4 p.u.,
as shown in the third and the fourth rows of \figurename{\ref{fig:exp_phase_jump_neg60}}.

Similarly, in \figurename{\ref{fig:exp_phase_jump_pos60}},
the grid-voltage angle suddenly jumps by $+$60$^\circ$ at $t=0$\,s.
The proposed control structure
effectively limits the transient overcurrent by 0.6 p.u.
compared to the conventional DSDU control structure.

It is worth noting that 
the `RST' signal from the CLA
may be triggered multiple times
during the transient period,
as shown in the last rows of 
\figurename{\ref{fig:exp_sym_voltage_sag_start_zoom}},
\ref{fig:exp_sym_voltage_sag_end_zoom},
\ref{fig:exp_l2l_voltage_sag_start_zoom},
\ref{fig:exp_l2l_voltage_sag_end_zoom},
\ref{fig:exp_phase_jump_neg60}, and
\ref{fig:exp_phase_jump_pos60}.
This is because PCC 
voltage fluctuates during the transient period after the grid fault happens
rather than immediately settling to a steady-state value.
Due to the interactions between the 
capacitors of the LC filter and the grid impedance,
the PCC voltage may exhibit short-term high-frequency oscillations
before settling down.
Multiple `RST' signals imply that
the control structure is responding to these fluctuations
to adjust the PWM carrier phase.
Although 
this may introduce additional switching events,
the increase in switching frequency is only temporary.
This is because the 
high-frequency oscillations are rapidly attenuated 
by the damping effect of the grid impedance,
particularly due to the skin effect at high frequencies \cite{Velaga2020HighFrequencySignature,Felipe2021SurveyofTraveling}.
Consequently,
this temporary increase in switching actions
does not significantly affect the overall switching loss
nor does it compromise the thermal performance of the inverter.

\section{Conclusion}\label{sec:conclusion}
This paper proposed a sub-switching-period current limiting strategy 
to effectively mitigate instantaneous overcurrents in grid-connected inverters.
The key contribution lies in its software-based implementation 
using a commonly used DSP without requiring auxiliary circuits or FPGAs.
By leveraging a multi-rate control structure and 
immediately shifting the PWM carrier upon fault detection, 
the proposed method overcomes the inherent delay 
limitations of conventional digital control, 
ensuring robust current limitation even at low switching frequencies.
Crucially, this is achieved without interrupting the current flow 
or disabling PWM pulses, thereby maintaining continuous grid support.
Experimental results under various grid faults validated 
that the proposed strategy effectively
limited transient overcurrents to within 1.0 p.u., 
whereas the conventional digital control structure exhibited 
overshoots of up to 1.5 p.u. 
Consequently, the proposed control 
structure offers a practical and cost-effective 
solution for enhancing the fault-ride-through 
capability and reliability of grid-tied inverters.





{
\vspace{\baselineskip}
\bibliographystyle{IEEEtran}
\bibliography{bibligraphy}

@ARTICLE{IEEESTD1547_2018,
  author={},
  journal={IEEE Std 1547-2018 (Revision of IEEE Std 1547-2003)}, 
  title={IEEE Standard for Interconnection and Interoperability of Distributed Energy Resources with Associated Electric Power Systems Interfaces}, 
  year={2018},
  volume={},
  number={},
  pages={1-138},
  doi={10.1109/IEEESTD.2018.8332112}}

@techreport{ramirez2024review,
  title={Review of Technical Requirements for Inverter-Based Resources in Chile},
  author={Ramirez, Lina and Ross, Haley and Velar, V{\'\i}ctor and Quintana, Eugenio and Veloso, Sim{\'o}n and Peralta, Jaime},
  year={2024},
  institution={National Renewable Energy Laboratory (NREL), Golden, CO (United States)}
}

@INPROCEEDINGS{Liu2009Triple,
  author={Liu, Dehong and Hu, An and Wang, Guangsen and Guo, Junhua},
  booktitle={2009 International Conference on Applied Superconductivity and Electromagnetic Devices}, 
  title={Triple-loop-controlled overload or short circuit current limiter and protection for three phase inverter}, 
  year={2009},
  volume={},
  number={},
  pages={201-205},
  doi={10.1109/ASEMD.2009.5306657}}

@ARTICLE{Paquette2015VirtualImpedanceCurrent,
  author={Paquette, Andrew D. and Divan, Deepak M.},
  journal={IEEE Transactions on Industry Applications}, 
  title={Virtual Impedance Current Limiting for Inverters in Microgrids With Synchronous Generators}, 
  year={2015},
  volume={51},
  number={2},
  pages={1630-1638},
  doi={10.1109/TIA.2014.2345877}}

@techreport{Lara2025April28th2025,
title = "April 28th 2025 Iberian Blackout: Analysis of Available Information",
author = "Lara, \{Jose Daniel\} and Ben Kroposki and Tarek Elgindy and Rodrigo Henriquez-Auba and Jarrad Wright",
year = "2025",
doi = "10.2172/2587951",
language = "American English",
type = "Other",
institution={National Renewable Energy Laboratory (NREL), Golden, CO (United States)},
howpublished = {\url{https://www.nrel.gov/docs/fy25osti/95103.pdf}},
}

@INPROCEEDINGS{Li2016InrushTransientCurrent,
  author={Li, Zhongyu and Zhao, Rende and Xin, Zhen and Guerrero, Josep M. and Savaghebi, Mehdi and Li, Peide},
  booktitle={2016 IEEE Applied Power Electronics Conference and Exposition (APEC)}, 
  title={Inrush Transient Current analysis and suppression of photovoltaic grid-connected inverters during voltage sag}, 
  year={2016},
  volume={},
  number={},
  pages={3697-3703},
  doi={10.1109/APEC.2016.7468402}}

@ARTICLE{He2022ReviewOfMultisampling,
  author={He, Shan and Zhou, Dao and Wang, Xiongfei and Zhao, Zhaoyang and Blaabjerg, Frede},
  journal={IEEE Transactions on Power Electronics}, 
  title={A Review of Multisampling Techniques in Power Electronics Applications}, 
  year={2022},
  volume={37},
  number={9},
  pages={10514-10533},
  doi={10.1109/TPEL.2022.3169662}}

@ARTICLE{Li2022AnInrushCurrent,
  author={Li, Haiguo and Gao, Zihan and Ji, Shiqi and Ma, Yiwei and Wang, Fei},
  journal={IEEE Journal of Emerging and Selected Topics in Power Electronics}, 
  title={An Inrush Current Limiting Method for Grid-Connected Converters Considering Grid Voltage Disturbances}, 
  year={2022},
  volume={10},
  number={2},
  pages={2608-2618},
  doi={10.1109/JESTPE.2022.3147515}}

@ARTICLE{Nagai2018ZVRTCapability,
  author={Nagai, Satoshi and Kusaka, Keisuke and Itoh, Jun-ichi},
  journal={IEEE Transactions on Industry Applications}, 
  title={ZVRT Capability of Single-Phase Grid-Connected Inverter With High-Speed Gate-Block and Minimized  LCL Filter Design}, 
  year={2018},
  volume={54},
  number={5},
  pages={5387-5399},
  doi={10.1109/TIA.2018.2858560}}

@INPROCEEDINGS{Wu2015AComprehensiveInvestigation,
  author={Wu, Rui and Diaz Reigosa, Paula and Iannuzzo, Francesco and Wang, Huai and Blaabjerg, Frede},
  booktitle={2015 17th European Conference on Power Electronics and Applications (EPE'15 ECCE-Europe)}, 
  title={A comprehensive investigation on the short circuit performance of MW-level IGBT power modules}, 
  year={2015},
  volume={},
  number={},
  pages={1-9},
  doi={10.1109/EPE.2015.7311761}}

@ARTICLE{Mohiuddin2025ATwoStageCurrent,
  author={Mohiuddin, Sheik M. and Du, Wei and Lyu, Xue and Kim, Jinho and Nguyen, Quan H.},
  journal={IEEE Transactions on Industry Applications}, 
  title={A Two-Stage Current Limiting Control Strategy for Direct-Droop-Controlled Grid-Forming Inverters}, 
  year={2025},
  volume={61},
  number={4},
  pages={6472-6483},
  doi={10.1109/TIA.2025.3544164}}

@ARTICLE{Wu2025ATransientCurrentLimiting,
  author={Wu, Mengze and Zhang, Xing and Hu, Kai and Zhao, Tao and Fu, Xinxin and Zhan, Xiangdui and Du, Lei and Li, Ming},
  journal={IEEE Journal of Emerging and Selected Topics in Power Electronics}, 
  title={A Transient Current-Limiting Control for the Grid-Forming (GFM) Cascaded H-Bridge (CHB) Converter}, 
  year={2025},
  volume={13},
  number={5},
  pages={6225-6240},
  doi={10.1109/JESTPE.2025.3602412}}

@ARTICLE{Baeckeland2024OvercurrentLimitinginGridForming,
  author={Baeckeland, Nathan and Chatterjee, Debjyoti and Lu, Minghui and Johnson, Brian and Seo, Gab-Su},
  journal={IEEE Transactions on Power Electronics}, 
  title={Overcurrent Limiting in Grid-Forming Inverters: A Comprehensive Review and Discussion}, 
  year={2024},
  volume={39},
  number={11},
  pages={14493-14517},
  doi={10.1109/TPEL.2024.3430316}}

@ARTICLE{Dong2023APLL_lessVoltage,
  author={Dong, Xiaofeng and Li, Hui},
  journal={IEEE Transactions on Power Electronics}, 
  title={A PLL-Less Voltage Sensorless Direct Deadbeat Control for a SiC Grid-Tied Inverter With LVRT Capability Under Wide-Range Grid Impedance}, 
  year={2023},
  volume={38},
  number={8},
  pages={9469-9481},
  doi={10.1109/TPEL.2023.3271202}}

@INPROCEEDINGS{Kawashima2024UltraRobust1MHz,
  author={Kawashima, Kaya and Kubo, Shinnosuke and Kreszens, Hidayat and Seki, Kohsuke and Yamabe, Kenta and Yokoyama, Tomoki},
  booktitle={2024 IEEE Energy Conversion Congress and Exposition (ECCE)}, 
  title={Ultra Robust 1MHz Multi-sampling Deadbeat Control for Megawatt-Class Grid-tied Multi-level Inverter with LC type Output Filter Configuration using C-HIL}, 
  year={2024},
  volume={},
  number={},
  pages={204-209},
  doi={10.1109/ECCE55643.2024.10861712}}

@ARTICLE{Tian2020MultirateHarmonic,
  author={Tian, Hao and Li, Yun Wei and Zhao, Qing},
  journal={IEEE Transactions on Power Electronics}, 
  title={Multirate Harmonic Compensation Control for Low Switching Frequency Converters: Scheme, Modeling, and Analysis}, 
  year={2020},
  volume={35},
  number={4},
  pages={4143-4156},
  doi={10.1109/TPEL.2019.2933770}}

@article{Kim1996ANovelVoltage,
  title={A Novel Voltage Modulation Technique of the Space Vector PWM},
  author={Joohn-Sheok Kim and Seung-Ki Sul},
  journal={IEEJ Transactions on Industry Applications},
  volume={116},
  number={8},
  pages={820-825},
  year={1996},
  doi={10.1541/ieejias.116.820}
}

@ARTICLE{Chung1998UnifiedVoltage,
  author={Dae-Woong Chung and Joohn-Sheok Kim and Sul, Seung-Ki},
  journal={IEEE Transactions on Industry Applications}, 
  title={Unified voltage modulation technique for real-time three-phase power conversion}, 
  year={1998},
  volume={34},
  number={2},
  pages={374-380},
  doi={10.1109/28.663482}}

@INPROCEEDINGS{Yu2017ImplementationofMultiSampling,
  author={Changzhou, Yu and Chun, Liu and Qionglong, Wang and Weitang, Zhang and Sicong, Li and Xing, Zhang},
  booktitle={2017 32nd Youth Academic Annual Conference of Chinese Association of Automation (YAC)}, 
  title={Implementation of multi-sampling current control for grid-connected inverters using TI TMS320F28377x}, 
  year={2017},
  volume={},
  number={},
  pages={1228-1233},
  doi={10.1109/YAC.2017.7967600}}

@article{launchxlF28379d,
	title = {{LAUNCHXL}-{F28379D} development kit},
	url = {https://www.ti.com/tool/LAUNCHXL-F28379D#tech-docs},
	urldate = {2026-01-03},
	journal = {Texas Instruments},
	note = {Accessed: Jan. 3, 2026},
}

@article{instruments2024tms320f2837xd,
  title={{TMS320F2837xD} dual-core real-time microcontrollers},
  author={},
  journal={Texas Instruments},
  year={2024},
  urldate = {2025-11-25},
  url={https://www.ti.com/lit/ds/sprs880p/sprs880p.pdf?ts=1767410186810},
  note={Accessed: Nov. 25, 2025},
}

@INPROCEEDINGS{Velaga2020HighFrequencySignature,
  author={Velaga, Yaswanth Nag and Prabakar, Kumaraguru and Singh, Akanksha and Sen, Pankaj K.},
  booktitle={2020 IEEE/IAS 56th Industrial and Commercial Power Systems Technical Conference}, 
  title={High-Frequency Signature-Based Fault Detection for Future MV Distribution Grids}, 
  year={2020},
  volume={},
  number={},
  pages={1-8},
  doi={10.1109/ICPS48389.2020.9176747}}

@ARTICLE{Felipe2021SurveyofTraveling,
  author={Wilches-Bernal, Felipe and Bidram, Ali and Reno, Matthew J. and Hernandez-Alvidrez, Javier and Barba, Pedro and Reimer, Benjamin and Montoya, Rudy and Carr, Christopher and Lavrova, Olga},
  journal={IEEE Access}, 
  title={A Survey of Traveling Wave Protection Schemes in Electric Power Systems}, 
  year={2021},
  volume={9},
  number={},
  pages={72949-72969},
  doi={10.1109/ACCESS.2021.3080234}}
}

\end{document}